\documentclass[nofootinbib, pra,  twocolumn,superscriptaddress]{revtex4-2}

\usepackage[T1]{fontenc}
\usepackage[latin9]{inputenc}
\usepackage{color}
\usepackage{amsmath}
\usepackage{amssymb}
\usepackage{graphicx}
\usepackage[unicode=true,pdfusetitle,
 bookmarks=true,bookmarksnumbered=false,bookmarksopen=false,
 breaklinks=false,pdfborder={0 0 0},pdfborderstyle={},backref=false,colorlinks=true]
 {hyperref}
\hypersetup{
 citecolor=blue,urlcolor=blue}

\makeatletter
\usepackage{bbm}

\makeatother

\begin{document}
\title{The field induced magnetic dipolar interaction for general boundary
conditions}
\author{Sheng-Wen Li}
\address{Center for Quantum Technology Research, and Key Laboratory of Advanced
Optoelectronic Quantum Architecture and Measurements, School of Physics,
Beijing Institute of Technology, Beijing 100081, China}
\author{Li-Ping Yang}
\affiliation{Center for Quantum Sciences and School of Physics, Northeast Normal
University, Changchun 130024, China}
\date{\today}
\begin{abstract}
By properly considering the propagation dynamics of the dipole field,
we obtain the full magnetic dipolar interaction between two quantum
dipoles for general situations. With the help the Maxwell equation
and the corresponding Green function, this result applies for general
boundary conditions, and naturally unifies all the interaction terms
between permanent dipoles, resonant or non-resonant transition dipoles,
and even the counter-rotating interaction terms altogether. In particular,
we study the dipolar interaction in a rectangular 3D cavity with discrete
field modes. When the two dipoles are quite near to each other and
far from the cavity boundary, their interaction simply returns the
freespace result; when the distance between the two dipoles is comparable
to their distance to the cavity boundary and the field mode wavelength,
the dipole images and near-resonant cavity modes bring in significant
changes to the freespace interaction. This approach also provides
a general way to study the interaction mediated by other kinds of
fields.
\end{abstract}
\maketitle

\section{Introduction}

The electric/magnetic dipolar interaction widely exists in many microscopic
systems, such as the Josephson qubit interacting with the dielectric
defects \citep{martinis_decoherence_2005,paik_observation_2011,rigetti_superconducting_2012,lisenfeld_decoherence_2016},
the nitrogen-vacancy (NV) interacting with the nuclear spins around
\citep{doherty_nitrogen-vacancy_2013,zhao_decoherence_2012}, as well
as the dipolar interactions in many chemical and biology molecular
\citep{yang_dimerization-assisted_2010,el-ganainy_resonant_2013,dong_photon-blockade_2016}.
In principle, the electromagnetic interactions between particles are
indirectly induced by their local interaction with the field. Thus,
it is possible to engineer the dipolar interaction by properly controlling
the mediating EM field \citep{lambert_cavity-mediated_2016,baranov_magnetic_2016,liao_dynamical_2016,shahmoon_dispersion_2013,donaire_dipole-dipole_2017,cortes_super-coulombic_2017,yang_single_2020,gonzalez-tudela_entanglement_2011,ying_extended_2019}.

But in literatures, the magnetic dipolar interaction between two quantum
dipoles with distance $R$ has three different descriptions:

1. \emph{Direct exchange}: In freespace, the classical dipolar interaction
between two static magnetic dipoles is ($\mathbf{e}_{\text{\textsc{r}}}$
is the unit directional vector of the distance $R$) \citep{jackson_classical_1998}
\begin{equation}
V_{0}=\frac{\mu_{0}}{4\pi R^{3}}\big[\vec{m}_{1}\cdot\vec{m}_{2}-3(\vec{m}_{1}\cdot\mathbf{e}_{\text{\textsc{r}}})(\vec{m}_{2}\cdot\mathbf{e}_{\text{\textsc{r}}})\big],\label{eq:classical}
\end{equation}
thus the quantum dipolar interaction is usually obtained by simply
replacing the classical dipole moments with quantum operators $\hat{\mathfrak{\boldsymbol{\mathfrak{m}}}}_{i}$.
Here the mediation effect of the EM field is not explicit, and the
frequencies of the quantum dipoles do not appear \citep{wrachtrup_processing_2006,zhao_decoherence_2012,doherty_nitrogen-vacancy_2013}
.

2. \emph{Master equation correction}: The dynamics of two dipoles
weakly interacting with the EM field can be described by a Markovian
master equation, where the system Hamiltonian contains an interaction
correction $\hat{V}_{12}=V(\omega)\,\hat{\tau}_{1}^{+}\hat{\tau}_{2}^{-}+\text{h.c.}$,
where $\hat{\tau}_{1,2}^{\pm}$ are the transition operators, and
\citep{lehmberg_radiation_1970,agarwal_quantum_1974,ficek_entangled_2002,henriques_exciton-polariton_2021}
\begin{align}
V(\omega)= & \frac{\mu_{0}}{4\pi R^{3}}\Big\{\vec{m}_{1}\cdot\vec{m}_{2}\,\big[(1-\eta^{2})\cos\eta+\eta\sin\eta\big]\label{eq:V-ME}\\
- & 3(\vec{m}_{1}\cdot\mathbf{e}_{\text{\textsc{r}}})(\vec{m}_{2}\cdot\mathbf{e}_{\text{\textsc{r}}})\big[(1-\frac{\eta^{2}}{3})\cos\eta+\eta\sin\eta\big]\Big\}.\nonumber 
\end{align}
Here $\eta:=\omega R/c$, and $\omega$ is the transition frequency
of the two resonant dipoles. The interaction strength $V(\omega)$
exhibits an oscillating decay with the distance $R$, and returns
the above Eq.\,(\ref{eq:classical}) when $\omega R/c\rightarrow0$.
However, since the rotating-wave approximation (RWA) must be applied
in a tricky way when deriving this master equation, this approach
could only give the interaction term between two resonant transition
dipoles with equal frequencies, while the other interaction terms
cannot be obtained, e.g., those between non-resonant dipoles \citep{shatokhin_coherence_2018},
permanent dipoles, and the counter rotating terms. Besides, this approach
cannot be applied when the field modes are discrete, e.g., in an ideal
lossless cavity.

3. \emph{Mode elimination}: Considering the two dipoles (frequency
$\omega_{1,2}$) both interacting with one common field mode (frequency
$\nu$), e.g., $\hat{V}=\hat{a}^{\dagger}(g_{1}\hat{\tau}_{1}^{-}+g_{2}\hat{\tau}_{2}^{-})+\text{h.c.}$,
the mediating field mode can be eliminated by the Fr\"ohlish-Nakajima
transform \citep{frohlich_theory_1950,nakajima_perturbation_1955,lambert_cavity-mediated_2016,li_long-term_2014,goldstein_dipole-dipole_1997},
which gives $\hat{V}_{12}\simeq\tilde{\beta}\,\hat{\tau}_{1}^{+}\hat{\tau}_{2}^{-}+\text{h.c.}$,
with (see Appendix \ref{sec:The-mediated-interaction}) 
\begin{equation}
\tilde{\beta}=\frac{1}{2}\big[\frac{g_{1}g_{2}^{*}}{\omega_{1}-\nu}+\frac{g_{2}g_{1}^{*}}{\omega_{2}-\nu}\big].\label{eq:adiabatic}
\end{equation}
In more realistic cases, usually more field modes should be involved.

These approaches are not always equivalent to each other, with different
application conditions, and it is not quite clear to see how these
approaches are connected with each other. In this paper, we make a
general approach which unifies all the above results together in the
same framework. The existence of one dipole would generate a dipole
field propagating to the other one, and then the dipolar interaction
is generated \citep{wang_magnetic_2018,hu_field-induced_2020}. The
dynamics of this dipole field is given by the Maxwell equation and
the corresponding Green function \citep{tai_dyadic_1994,donaire_dipole-dipole_2017,yang_single_2020,cortes_super-coulombic_2017,dung_resonant_2002}.
Based on this idea, we obtained the full dipolar interaction for two
quantum magnetic dipoles for general boundary conditions. Our result
naturally includes all the interaction terms between permanent dipoles,
resonant or non-resonant transition dipoles, and even the counter-rotating
interaction terms altogether \citep{hu_field-induced_2020}. These
terms are crucial for the delicate control in microscopic systems
under proper driving field, especially for magnetic dipolar interactions,
since one magnetic dipole operator usually contains both nonzero transition
and permanent dipole moments. Under proper conditions, our result
well reduces to all the above three cases.

In particular, we study the dipolar interaction between two magnetic
dipoles inside a rectangular 3D cavity made of ideal conductors, where
the field modes are fully discrete. It turns out, when the two dipoles
are quite near to each other and far from the boundaries, the interaction
always returns to the static dipolar interaction (\ref{eq:classical})
in freespace; when the dipoles are close to the cavity boundary, the
dipole field propagation is strongly restricted due to the conductor
boundary, which further influences the dipolar interaction generated. 

The paper is arranged as follows. In Sec. II, we show the derivation
for the dipolar interaction for general situations. In Sec. III, we
study how the Green function is evaluated in a rectangular 3D cavity.
In Sec. IV, we show the numerical results for the dipolar interaction.
In Sec. V, we discuss the possible application of these results in
different physical systems.

\section{Dipole field propagation}

We consider two magnetic dipoles are placed in the EM field. The Hamiltonian
of each dipole is $\hat{H}_{\alpha}=\sum_{u}\mathrm{E}_{u}^{(\alpha)}|u\rangle_{\alpha}\langle u|$,
where $\mathrm{E}_{u}^{(\alpha)}$and $|u\rangle_{\alpha}$ are the
eigen energy and the corresponding eigenstate of dipole-$\alpha$
($\alpha=1,2$).

The two magnetic dipoles interacts with the EM field via their local
interactions $\hat{V}_{\alpha}=-\hat{\boldsymbol{\mathfrak{m}}}_{\alpha}\cdot\hat{\mathbf{B}}(\mathbf{r}_{\alpha})$,
where $\mathbf{r}_{\alpha}$ is the position of dipole-$\alpha$,
and $\hat{\boldsymbol{\mathfrak{m}}}_{\alpha}:=\sum_{uv}\vec{m}_{\alpha}^{uv}\hat{\tau}_{\alpha}^{uv}$
is the dipole operator, with $\vec{m}_{\alpha}^{uv}:=\langle u|\hat{\boldsymbol{\mathfrak{m}}}_{\alpha}|v\rangle_{\alpha}$
and $\hat{\mathfrak{\tau}}_{\alpha}^{uv}:=|u\rangle_{\alpha}\langle v|$
(see Fig.\,\ref{fig-demo}). Usually, one dipole operator $\hat{\boldsymbol{\mathfrak{m}}}_{\alpha}$
contains both nonzero permanent dipoles (the diagonal terms $\vec{m}_{\alpha}^{uu}|u\rangle_{\alpha}\langle u|$)
and transition dipoles (the off-diagonal terms $\vec{m}_{\alpha}^{uv}|u\rangle_{\alpha}\langle v|$
with $u\neq v$) together \citep{wang_magnetic_2018}. Hereafter,
these vector operators are denoted as $\vec{m}_{\alpha}^{uv}|u\rangle_{\alpha}\langle v|:=\hat{\boldsymbol{\mathfrak{m}}}_{\alpha}^{uv}$,
and thus $\hat{\boldsymbol{\mathfrak{m}}}_{\alpha}=\sum_{uv}\hat{\boldsymbol{\mathfrak{m}}}_{\alpha}^{uv}$.

The existence of one magnetic dipole changes the EM field dynamics,
and when such field changes propagate to the other dipole, the interaction
is generated between the two dipoles \citep{wang_magnetic_2018,hu_field-induced_2020}.
Here we first consider the dipole field generated by dipole-1 interacting
with dipole-2. Notice that, the quantized magnetic field $\hat{\mathbf{B}}(\mathbf{r},t)$
also follows the Maxwell equation (Appendix \ref{sec:The-dynamical})
\begin{equation}
\big[\tfrac{1}{c^{2}}\partial_{t}^{2}-\nabla^{2}\big]\hat{\mathbf{B}}(\mathbf{r},t)=\mu_{0}\nabla\times\hat{\mathbf{J}}_{1}(\mathbf{r},t),\label{eq:maxwell}
\end{equation}
which has the same form with the classical one \citep{jackson_classical_1998,huttner_quantization_1992,scheel_qed_1998,scheel_quantum_2006,vogel_quantum_2006},
although the explicit form of the quantized field $\hat{\mathbf{B}}(\mathbf{r},t)$
is not written down. Here $\hat{\mathbf{J}}_{1}:=\nabla\times\hat{\mathbf{M}}_{1}$
is the electric current density induced by dipole-1, and $\hat{\mathbf{M}}_{1}(\mathbf{r},t):=\hat{\boldsymbol{\mathfrak{m}}}_{1}(t)\,\delta(\mathbf{r}-\mathbf{r}_{1})$
is the magnetization density. 

The dynamics of the quantized field contains two contributions $\hat{\mathbf{B}}(\mathbf{r},t)=\hat{\mathbf{B}}_{0}(\mathbf{r},t)+\hat{\mathbf{B}}_{\text{d1}}(\mathbf{r},t)$,
where $\hat{\mathbf{B}}_{0}(\mathbf{r},t)$ comes from the vacuum
EM field, given by the homogenous equation $\big[c^{-2}\partial_{t}^{2}-\nabla^{2}\big]\hat{\mathbf{B}}_{0}=0$;
$\hat{\mathbf{B}}_{\text{d1}}(\mathbf{r},t)$ is the dipole field
generated by dipole-1, which can be given with the help of the tensor
Green function \citep{tai_dyadic_1994,park_accelerated_2009,sanamzadeh_fast_2019}
\begin{gather}
\hat{\mathbf{B}}_{\text{d1}}(\mathbf{r},t)=\mu_{0}\int_{-\infty}^{\infty}dt'\int_{V}d^{3}r'\,\mathbb{G}_{\text{m}}(\mathbf{r}t,\mathbf{r}'t')\cdot\hat{\mathbf{J}}_{1}(\mathbf{r}',t'),\nonumber \\
\big[\tfrac{1}{c^{2}}\partial_{t}^{2}-\nabla^{2}\big]\mathbb{G}_{\text{m}}(\mathbf{r}t,\mathbf{r}'t')=\nabla\times\mathbb{I}\,\delta(\mathbf{r}-\mathbf{r}')\delta(t-t').\label{eq:dipole-field}
\end{gather}
Here $\mathbf{r}\,(\mathbf{r}')$ denotes the field (source) position
in the Green function $\mathbb{G}_{\text{m}}(\mathbf{r}t,\mathbf{r}'t')$.

In the above dipole field $\hat{\mathbf{B}}_{\text{d1}}(\mathbf{r},t)$,
the dynamics of dipole-1 is contained in the current $\hat{\mathbf{J}}_{1}(\mathbf{r},t)=\nabla\times[\hat{\boldsymbol{\mathfrak{m}}}_{1}(t)\delta(\mathbf{r}-\mathbf{r}_{1})]$.
Here we make an approximation that during the field propagation time
($\sim R/c$), the dynamics of the dipole-1 can be regarded as only
governed by its self-Hamiltonian $\hat{H}_{1}$ and thus follows the
unitary evolution, which gives $\hat{\boldsymbol{\mathfrak{m}}}_{\alpha}^{ab}(t')\simeq\hat{\boldsymbol{\mathfrak{m}}}_{\alpha}^{ab}(t)\,\exp[-i\omega_{ab}^{(\alpha)}(t'-t)]$,
with $\hbar\omega_{ab}^{(\alpha)}:=\mathrm{E}_{a}^{(\alpha)}-\mathrm{E}_{b}^{(\alpha)}$
\citep{wang_magnetic_2018,hu_field-induced_2020}. In most microscopic
experiments, the distance between different dipoles are within several
microns, thus the propagation time for their interactions ($R/c\lesssim10^{-15}\text{ s}$)
is much shorter than the decay time of each single dipole, which guarantees
this approximation reliable. Then the dipole field (\ref{eq:dipole-field})
can be further obtained as 

\begin{widetext}

\begin{align}
\hat{\mathbf{B}}_{\text{d1}}(\mathbf{r},t) & =\mu_{0}\int_{-\infty}^{\infty}dt'\int_{V}d^{3}r'\,\tilde{\mathbb{G}}_{\text{m}}(\mathbf{r}t,\mathbf{r}'t')\cdot\nabla'\times[\hat{\boldsymbol{\mathfrak{m}}}_{1}(t')\delta(\mathbf{r}'-\mathbf{r}_{1})]\nonumber \\
 & \simeq-\mu_{0}\sum_{ab}\int_{-\infty}^{\infty}dt'\int_{V}d^{3}r'\,\tilde{\mathbb{G}}_{\text{m}}(\mathbf{r}t,\mathbf{r}'t')\cdot\nabla_{1}\times[\hat{\boldsymbol{\mathfrak{m}}}_{1}^{ab}(t)e^{-\mathrm{i}\omega_{ab}^{(1)}(t'-t)}\delta(\mathbf{r}'-\mathbf{r}_{1})]\nonumber \\
 & =\mu_{0}\sum_{ab}\tilde{\mathbb{G}}_{\text{m}}(\mathbf{r},\mathbf{r}_{1};\omega_{ab}^{(1)})\times\overleftarrow{\nabla}_{1}\cdot\hat{\boldsymbol{\mathfrak{m}}}_{1}^{ab}(t).
\end{align}
 \end{widetext} Here $[\mathbb{G}\times\overleftarrow{\nabla}]_{ij}:=\epsilon_{jpq}\partial_{q}\mathbb{G}_{ip}$
means the curl operation to the left, $\nabla'/\nabla_{1}$ is the
derivative respect to $\mathbf{r}'/\mathbf{r}_{1}$, and 
\begin{equation}
\tilde{\mathbb{G}}_{\text{m}}(\mathbf{r},\mathbf{r}';\omega):=\int_{-\infty}^{\infty}dt'\,\mathbb{G}_{\text{m}}(\mathbf{r}t,\mathbf{r}'t')e^{-\mathrm{i}\omega(t'-t)}
\end{equation}
 is just the Fourier transform of the Green function $\mathbb{G}_{\text{m}}(\mathbf{r}t,\mathbf{r}'t')$.

Therefore, the local interaction between dipole-2 (at position \textbf{$\mathbf{r}_{2}$})
and the dipole field $\hat{\mathbf{B}}_{\text{d1}}(\mathbf{r},t)$
naturally gives the dipolar interaction as the following symmetric
form 
\begin{align}
\hat{V}_{2\leftarrow1} & =-\hat{\boldsymbol{\mathfrak{m}}}_{2}(t)\cdot\hat{\mathbf{B}}_{\text{d1}}(\mathbf{r}_{2},t)\label{eq:V21}\\
 & =\mu_{0}\sum_{ab,uv}\hat{\boldsymbol{\mathfrak{m}}}_{2}^{uv}\cdot\tilde{\mathbb{G}}_{\text{m}}(\mathbf{r}_{2},\mathbf{r}_{1};\omega_{ab}^{(1)})\times\overleftarrow{\nabla}_{1}\cdot\hat{\boldsymbol{\mathfrak{m}}}_{1}^{ab}\nonumber \\
 & =\mu_{0}\sum_{ab,uv}\hat{\boldsymbol{\mathfrak{m}}}_{2}^{uv}\cdot\nabla_{2}\times\tilde{\mathbb{G}}_{A}(\mathbf{r}_{2},\mathbf{r}_{1};\omega_{ab}^{(1)})\times\overleftarrow{\nabla}_{1}\cdot\hat{\boldsymbol{\mathfrak{m}}}_{1}^{ab}.\nonumber 
\end{align}
 Here $\tilde{\mathbb{G}}_{A}(\mathbf{r},\mathbf{r}';\omega)$ is
introduced from $\tilde{\mathbb{G}}_{\text{m}}(\mathbf{r},\mathbf{r}';\omega)=\nabla\times\tilde{\mathbb{G}}_{A}(\mathbf{r},\mathbf{r}';\omega)$,
which is just the Green function for the vector potential $\mathbf{B}=\nabla\times\mathbf{A}$.
And $\tilde{\mathbb{G}}_{A}(\mathbf{r},\mathbf{r}';\omega)$ follows
\begin{equation}
[\nabla^{2}+(\omega/c)^{2}]\tilde{\mathbb{G}}_{A}(\mathbf{r},\mathbf{r}';\omega)=-\mathbb{I}\,\delta(\mathbf{r}-\mathbf{r}').\label{eq:Green-eq}
\end{equation}
The interaction (\ref{eq:V21}) is gauge independent, since the term
$\nabla\times\tilde{\mathbb{G}}_{A}$ in $\hat{V}_{2\leftarrow1}$
remains unchanged in gauge transformations. Besides, it is worth noting
that the RWA is not needed throughout the above derivations.

Therefore, once $\tilde{\mathbb{G}}_{A}(\mathbf{r},\mathbf{r}';\omega)$
is solved, the dipolar interaction $\hat{V}_{2\leftarrow1}$ can be
obtained from Eq.\,(\ref{eq:V21}). The full dipolar interaction
between the two magnetic dipoles also should involve the interaction
between dipole-1 and the field induced from dipole-2, i.e.,
\begin{align}
\hat{V}_{1\leftrightarrow2} & =\frac{1}{2}\big[\hat{V}_{2\leftarrow1}+\hat{V}_{1\leftarrow2}\big]\nonumber \\
 & =-\frac{1}{2}\big[\hat{\boldsymbol{\mathfrak{m}}}_{2}\cdot\hat{\mathbf{B}}_{\text{d1}}(\mathbf{r}_{2})+\hat{\boldsymbol{\mathfrak{m}}}_{1}\cdot\hat{\mathbf{B}}_{\text{d2}}(\mathbf{r}_{1})\big].\label{eq:Vint}
\end{align}

In freespace, the Green equation (\ref{eq:Green-eq}) has an isotropic
solution $\tilde{\mathbb{G}}_{A}(\mathbf{r}_{2},\mathbf{r}_{1};\,\omega\equiv ck)=(\cos kR/4\pi R)\,\mathbb{I}$,
with $R\equiv|\mathbf{r}_{2}-\mathbf{r}_{1}|$. Then Eq.\,(\ref{eq:V21})
gives the dipolar interaction as $\hat{V}_{2\leftarrow1}=\sum_{uv,ab}V_{2\leftarrow1}^{uv,ab}(\omega_{ab}^{(1)})\,\hat{\tau}_{2}^{uv}\hat{\tau}_{1}^{ab}$,
where the interaction strength $V_{2\leftarrow1}^{uv,ab}(\omega_{ab}^{(1)})$
of each term is given by \begin{widetext}
\begin{align}
V_{2\leftarrow1}^{uv,ab}(\omega) & =\mu_{0}\Big[\vec{m}_{2}^{uv}\cdot\vec{m}_{1}^{ab}\nabla^{2}-(\vec{m}_{2}^{uv}\cdot\nabla)(\vec{m}_{1}^{ab}\cdot\nabla)\Big]\frac{\cos kR}{4\pi R}\nonumber \\
 & =\frac{\mu_{0}}{4\pi R^{3}}\Big\{\vec{m}_{2}^{uv}\cdot\vec{m}_{1}^{ab}\,\big[(1-\eta^{2})\cos\eta+\eta\sin\eta\big]-3(\vec{m}_{2}^{uv}\cdot\mathbf{e}_{\text{\textsc{r}}})(\vec{m}_{1}^{ab}\cdot\mathbf{e}_{\text{\textsc{r}}})\big[(1-\frac{\eta^{2}}{3})\cos\eta+\eta\sin\eta\big]\Big\},\label{eq:V21_uv,ab}
\end{align}
\end{widetext} where $\eta=\omega R/c=kR$. This result just has
the same form as the master equation approach (\ref{eq:V-ME}), and
here $\hat{\boldsymbol{\mathfrak{m}}}_{2}^{uv}$ and $\hat{\boldsymbol{\mathfrak{m}}}_{1}^{ab}$
no longer need to be resonant dipoles. 

The different matrix elements of the operator $\hat{V}_{2\leftarrow1}$
indicates different kinds of interactions. For example, considering
each dipole only has two levels $|\mathrm{g}\rangle_{\alpha},|\mathrm{e}\rangle_{\alpha}$
as the ground and excited states, the interaction term $\langle\mathrm{e}_{1}\mathrm{g}_{2}|\hat{V}_{2\leftarrow1}|\mathrm{g}_{1}\mathrm{e}_{2}\rangle\,|\mathrm{e}_{1}\mathrm{g}_{2}\rangle\langle\mathrm{g}_{1}\mathrm{e}_{2}|\equiv V_{2\leftarrow1}^{\mathrm{ge},\mathrm{eg}}(\omega_{\mathrm{eg}}^{(1)})\,\hat{\tau}_{1}^{+}\hat{\tau}_{2}^{-}$
gives the interaction between two transition dipoles ($\hat{\tau}_{1}^{+}=|\mathrm{e}\rangle_{1}\langle\mathrm{g}|$
and $\hat{\tau}_{2}^{-}=|\mathrm{g}\rangle_{2}\langle\mathrm{e}|$),
which just returns Eq.\,(\ref{eq:V-ME}); the interaction term $\langle\mathrm{e}_{1}\mathrm{g}_{2}|\hat{V}_{2\leftarrow1}|\mathrm{e}_{1}\mathrm{g}_{2}\rangle\,|\mathrm{e}_{1}\mathrm{g}_{2}\rangle\langle\mathrm{e}_{1}\mathrm{g}_{2}|\equiv V_{2\leftarrow1}^{\mathrm{gg},\mathrm{ee}}(\omega_{\mathrm{ee}}^{(1)}=0)\,\hat{\tau}_{1}^{\mathrm{ee}}\hat{\tau}_{2}^{\mathrm{gg}}$
indicates the interaction between two permanent dipoles, namely, when
dipole-1 is in $|\mathrm{e}\rangle_{1}$ and dipole-2 is in $|\mathrm{g}\rangle_{2}$,
which exactly returns the static dipolar interaction (\ref{eq:classical}).
Besides, $\hat{V}_{2\leftarrow1}$ also includes the interaction terms
between one permanent dipole and one transition dipole, e.g., $V_{2\leftarrow1}^{\mathrm{ee},\mathrm{eg}}(\omega_{\mathrm{eg}}^{(1)})\,\hat{\tau}_{1}^{+}\hat{\tau}_{2}^{\mathrm{ee}}$,
and counter-rotating terms such as $V_{2\leftarrow1}^{\mathrm{eg},\mathrm{eg}}(\omega_{\mathrm{eg}}^{(1)})\,\hat{\tau}_{1}^{+}\hat{\tau}_{2}^{+}$
\citep{hu_field-induced_2020}. Though usually neglected, these counter-rotating
terms could exhibit significant physical effects under proper driving
field. 

On the other hand, the interaction propagated from dipole-2 is $\hat{V}_{1\leftarrow2}=\sum_{uv,ab}V_{1\leftarrow2}^{ab,uv}(\omega_{uv}^{(2)})\,\hat{\tau}_{1}^{ab}\hat{\tau}_{2}^{uv}$,
and the full interaction is $\hat{V}_{1\leftrightarrow2}=\frac{1}{2}(\hat{V}_{1\leftarrow2}+\hat{V}_{2\leftarrow1})$.
Because of the isotropy of $\tilde{\mathbb{G}}_{A}(|\mathbf{r}_{1}-\mathbf{r}_{2}|\equiv R)$
in freespace, we have $V_{1\leftarrow2}^{ab,uv}(\omega)=V_{2\leftarrow1}^{uv,ab}(\omega)$,
except now the frequency is $\omega_{uv}^{(2)}$ from dipole-2. For
the resonant case $|\omega_{ab}^{(1)}|=|\omega_{uv}^{(2)}|$, the
interaction strengths contributed from both $1\leftarrow2$ and $2\leftarrow1$
directions are equal.

It is also worth noticing that, the explicit form of the quantized
field $\hat{\mathbf{B}}(\mathbf{r},t)$ is not needed throughout the
above derivations, and the starting point is simply the Maxwell equation
(\ref{eq:maxwell}), thus the above results do not depend on how the
EM field is quantized (e.g., whether the Coulomb or Lorenz gauge is
used). The tensor Green function $\tilde{\mathbb{G}}_{A}(\mathbf{r},\mathbf{r}';\omega)$
is the same as the one in classical electrodynamics, and the results
here apply for general boundary conditions. 

\section{The dipolar interaction inside a lossless cavity}

Now we further consider the dipolar interaction between two dipoles
inside a rectangular cavity, which is made of ideal conductors with
no loss. 

In this case, the field modes in the cavity are fully discrete. The
above discussions about the dipole field propagation still holds,
and the cavity boundary condition is naturally included in the tensor
Green function $\tilde{\mathbb{G}}_{A}(\mathbf{r},\mathbf{r}';\omega)$
from Eq.\,(\ref{eq:Green-eq}), i.e.,
\begin{equation}
\hat{n}\times\tilde{\mathbb{G}}_{A}(\mathbf{r},\mathbf{r}';\omega)=0,\quad\nabla\cdot\tilde{\mathbb{G}}_{A}(\mathbf{r},\mathbf{r}';\omega)=0
\end{equation}
 for $\mathbf{r}$ on the conductor plane \citep{park_accelerated_2009,sanamzadeh_fast_2019}.
Here the Green function can be written as \textbf{$\tilde{\mathbb{G}}_{A}=G_{A}^{x}\,\mathbf{e}_{x}\mathbf{e}_{x}+G_{A}^{y}\,\mathbf{e}_{y}\mathbf{e}_{y}+G_{A}^{z}\,\mathbf{e}_{z}\mathbf{e}_{z}$},
and the above boundary condition indicates $G_{A}^{\sigma}=0$ on
the sidewalls and $\partial_{\sigma}G_{A}^{\sigma}=0$ on the end
caps with respect to direction-$\sigma$ (for $\sigma=x,y,z$). 

The Green function $\tilde{\mathbb{G}}_{A}(\mathbf{r},\mathbf{r}';\omega)$
has the following solution of mode expansion 
\begin{equation}
G_{A}^{\sigma}(\mathbf{r},\mathbf{r}';\omega)=\sum_{\mathbf{k}}\frac{\mathtt{A}_{\mathbf{k}}^{\sigma}(\mathbf{r})\mathtt{A}_{\mathbf{k}}^{\sigma}(\mathbf{r}')}{\mathbf{k}^{2}-(\omega/c)^{2}},\label{eq:mode-expansion}
\end{equation}
where $\big\{\vec{\mathtt{A}}_{\mathbf{k}}(\mathbf{r})\big\}$ is
a set of orthonormal eigenfunctions for $[\nabla^{2}+\mathbf{k}^{2}]\vec{\mathtt{A}}_{\mathbf{k}}(\mathbf{r})=0$,
with $\mathbf{k}$ indexing the field modes (see Appendix \ref{sec:Field-eigen-modes}).
Then the dipolar interaction can be further obtained {[}from Eq.\,(\ref{eq:V21}){]}
as $\hat{V}_{2\leftarrow1}=\sum_{uv,ab}V_{2\leftarrow1}^{uv,ab}(\omega_{ab}^{(1)})\,\hat{\tau}_{2}^{uv}\hat{\tau}_{1}^{ab}$,
and the interaction strength of each term is given by 
\begin{equation}
V_{2\leftarrow1}^{uv,ab}(\omega)=\sum_{\mathbf{k}}\frac{c^{2}\zeta_{2,\mathbf{k}}^{uv}\zeta_{1,\mathbf{k}}^{ab}}{\omega^{2}-c^{2}\mathbf{k}^{2}}\sim\sum_{|\mathbf{k}|\simeq\frac{\omega}{c}}\frac{c^{2}\zeta_{2,\mathbf{k}}^{uv}\zeta_{1,\mathbf{k}}^{ab}/2\omega}{\omega-c|\mathbf{k}|},\label{eq:Gk-expansion}
\end{equation}
where $\zeta_{\alpha,\mathbf{k}}^{ab}:=\sqrt{\mu_{0}}\,\vec{\mathfrak{m}}_{\alpha}^{ab}\cdot\nabla\times\vec{\mathtt{A}}_{\mathbf{k}}(\mathbf{r}_{\alpha})$
for $\alpha=1,2$.

Intuitively, since $\omega^{2}-c^{2}\mathbf{k}^{2}$ appears in the
denominator, we may expect that only the near-resonant terms with
$|\mathbf{k}|\simeq\omega/c$ dominates in the summation, and that
would give an interaction strength returning the result (\ref{eq:adiabatic})
in the mode elimination approach (see Appendix \ref{sec:The-mediated-interaction}).
However, it turns out the above summation series converges too slowly,
since the density of state of the field modes also increases as $\sim\mathbf{k}^{2}$,
which is in the same scale with the denominator $\omega^{2}-c^{2}\mathbf{k}^{2}$.
As a result, only counting the near-resonant terms is not enough to
give a precise evaluation for the dipolar interaction.

\begin{figure}
\begin{centering}
\includegraphics[width=0.98\columnwidth]{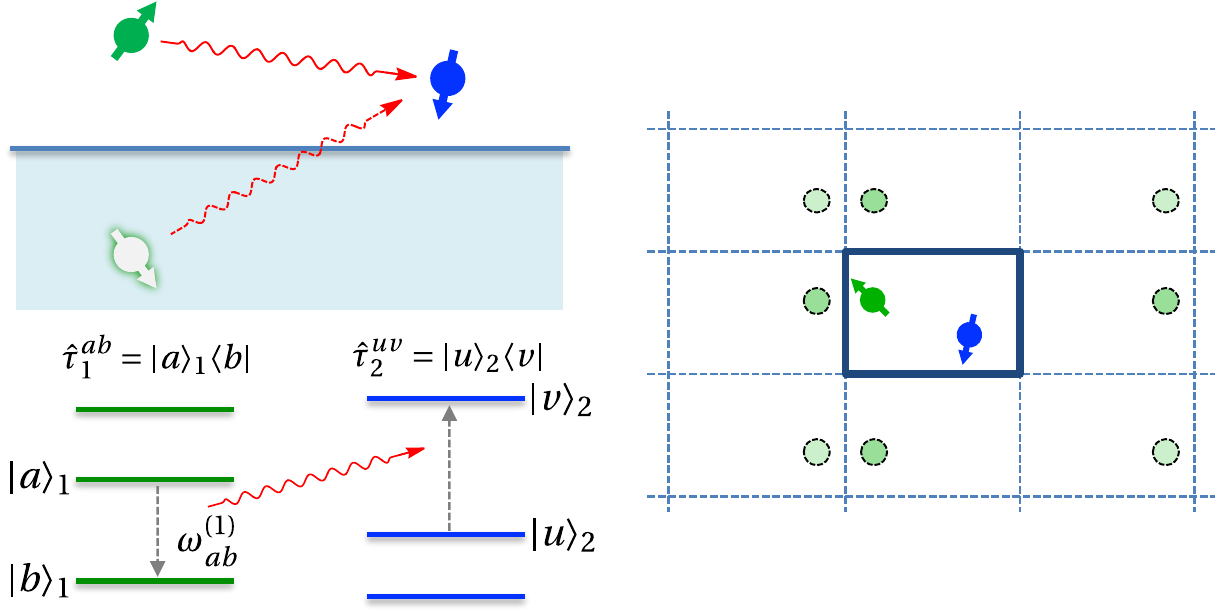}
\par\end{centering}
\caption{Demonstration for the dipole field propagation and the image sources.}
\label{fig-demo}
\end{figure}

On the other hand, $G_{A}^{\sigma}(\mathbf{r},\mathbf{r}';\omega\equiv ck)$
also can be written down in the form of image expansion (Fig.\,\ref{fig-demo}),
but that also has the slow converging problem for numerical estimations.
This problem can be solved by the improved Ewald expansion \citep{park_accelerated_2009,sanamzadeh_fast_2019},
that is, with the help of $\text{erf}(x)+\text{erfc}(x)=1$, the Green
function is separated as $G_{A}^{\sigma}(\mathbf{r},\mathbf{r}';\omega)\equiv G_{A1}^{\sigma}(\mathbf{r},\mathbf{r}';\omega)+G_{A2}^{\sigma}(\mathbf{r},\mathbf{r}';\omega)$,
which further gives (see more details in Refs.\,{[}\onlinecite{park_accelerated_2009}, \onlinecite{sanamzadeh_fast_2019}{]})
\begin{align}
G_{A1}^{\sigma} & =\sum_{ijl,rst}(-1)^{r+s+t-q_{\sigma}}\frac{\cos kR_{ijl,rst}}{4\pi R_{ijl,rst}}\cdot\text{erfc}(K_{c}R_{ijl,rst}),\nonumber \\
G_{A2}^{\sigma} & =\sum_{ijl,rst}(-1)^{r+s+t-q_{\sigma}}\frac{\cos kR_{ijl,rst}}{4\pi R_{ijl,rst}}\cdot\text{erf}(K_{c}R_{ijl,rst})\nonumber \\
 & =\sum_{\mathrm{npq}}\mathtt{A}_{\mathrm{npq}}^{\sigma}(\mathbf{r})\mathtt{A}_{\mathrm{npq}}^{\sigma}(\mathbf{r}')\cdot\Gamma^{(K_{c})}(k,k_{\text{\ensuremath{\mathrm{npq}}}}).\label{eq:G1-G2}
\end{align}
Here $\mathbf{k}\equiv(\mathrm{n}\pi/L_{x},\,\mathrm{p}\pi/L_{y},\,\mathrm{q}\pi/L_{z})$
is the field mode index with $k_{\mathrm{npq}}:=|\mathbf{k}|$ and
$\mathrm{n},\mathrm{p},\mathrm{q}\in\mathbb{Z}_{0}^{+}$. $R_{ijl,rst}:=|\mathbf{r}-\mathbf{R}_{ijl,rst}'|$
is the distance between the field point $\mathbf{r}$ and the image
of $\mathbf{r}'\equiv(x',y',z')$ at 
\begin{align}
\mathbf{R}_{ijl,rst}'= & [2iL_{x}+(-1)^{r}x']\,\mathbf{e}_{x}+[2jL_{y}+(-1)^{s}y']\,\mathbf{e}_{y}\nonumber \\
 & +[2lL_{z}+(-1)^{t}z']\,\mathbf{e}_{z},
\end{align}
 with $i,j,l\in\mathbb{Z}$, $r,s,t=0,1$, and $q_{x,y,z}=r,s,t$.
The function $\Gamma^{(K_{c})}(k,k_{\mathrm{npq}})$ provides a fast-converging
cutoff, 
\[
\Gamma^{(K_{c})}(k,k_{\mathrm{npq}})=\frac{1}{2k_{\mathrm{npq}}}\big[\frac{e^{-\frac{(k+k_{\mathrm{npq}})^{2}}{4K_{c}^{2}}}}{k_{\mathrm{npq}}+k}+\frac{e^{-\frac{(k-k_{\mathrm{npq}})^{2}}{4K_{c}^{2}}}}{k_{\mathrm{npq}}-k}\big].
\]
Here $K_{c}$ is a free parameter {[}usually set as $K_{c}=\sqrt{\pi}/(2V^{1/3})${]}.

Now both the above $G_{A1}^{\sigma}$ and $G_{A2}^{\sigma}$ converge
rapidly enough for numerical evaluations (in our numerical results
below, $\sim10^{3}$ image terms are counted, see also the precision
analysis in Ref. \citep{park_accelerated_2009,sanamzadeh_fast_2019}).
When $K_{c}\rightarrow0$, the Green function $\tilde{\mathbb{G}}_{A}(\mathbf{r},\mathbf{r}';\omega)$
gives the form of image expansion, and when $K_{c}\rightarrow\infty$,
$\tilde{\mathbb{G}}_{A}(\mathbf{r},\mathbf{r}';\omega)$ returns the
form of mode expansion (\ref{eq:mode-expansion}). In this sense,
we say $G_{A1}^{\sigma}$ and $G_{A2}^{\sigma}$ indicate the contributions
from the images and mode-propagation respectively. Then the dipolar
interaction strength can be further obtained from Eqs.\,(\ref{eq:V21},
\ref{eq:Vint}).

\section{Numerical results}

\begin{figure*}
\begin{centering}
\includegraphics[width=1\textwidth]{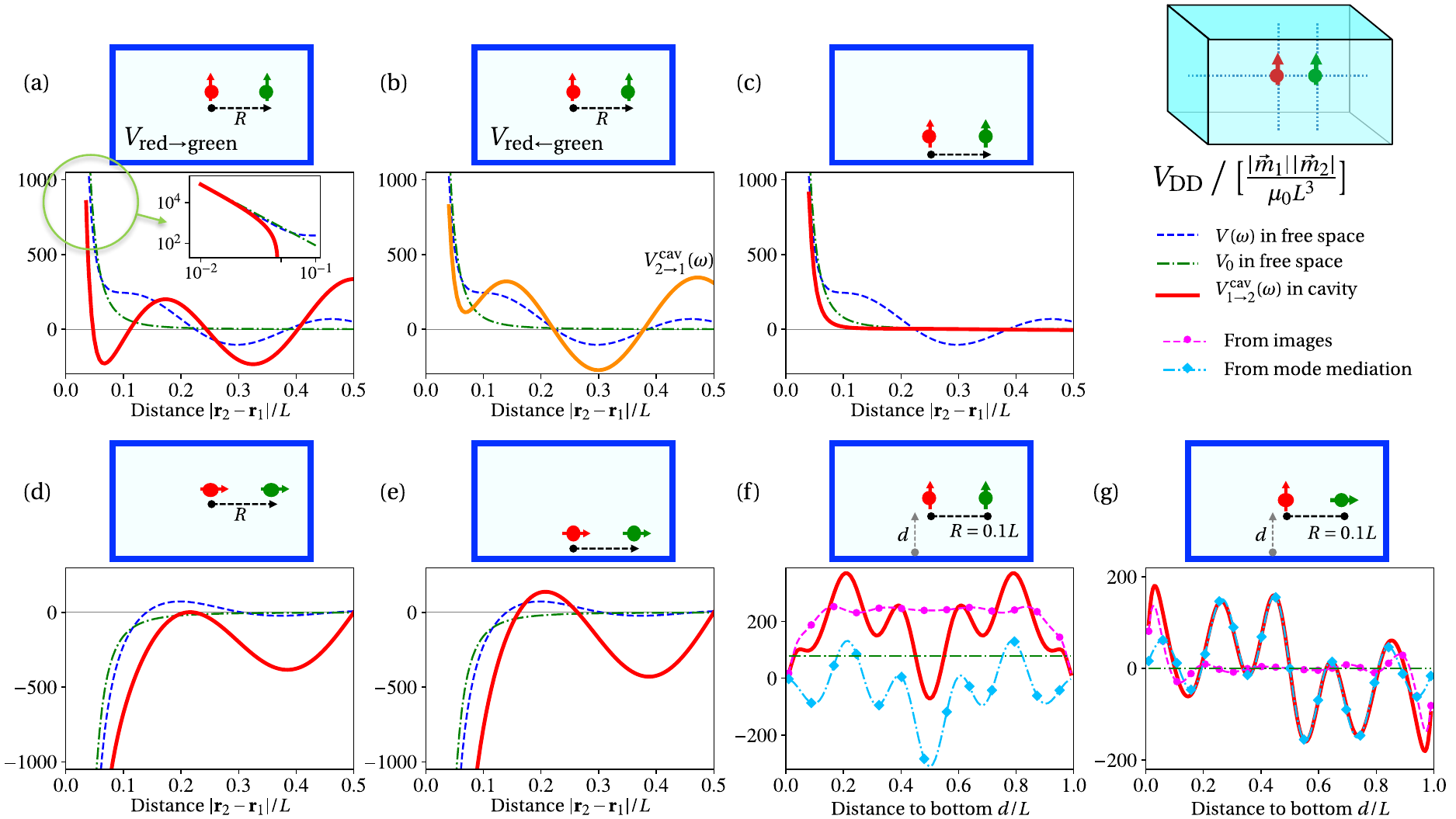}
\par\end{centering}
\caption{The dipolar interaction in a rectangular 3D cavity (solid red). (a)
Dipole-1 (red) is placed in the box center $\mathbf{r}_{1}=(0.5L_{x},\,0.5L_{y},\,0.5L_{z})$
with $L_{x,y,z}\equiv L$, and dipole-2 (green) is placed at $\mathbf{r}_{2}=\mathbf{r}_{1}+R\,\mathbf{e}_{x}$
as all the other figures. Both $\vec{m}_{1,2}$ are oriented at $z$-direction,
and the interaction $V_{2\leftarrow1}^{\text{cav}}(\omega)$ is generated
between dipole-2 and the field from dipole-1, with $\omega/c=20\,L^{-1}$.
(b) All the conditions are the same with (a) except the interaction
$V_{1\leftarrow2}^{\text{cav}}(\omega)$ is generated between dipole-1
and the field from dipole-2. (c) The two dipoles are placed near the
bottom $\mathbf{r}_{1}=(0.5L_{x},\,0.5L_{y},\,0.01L_{z})$. (d) Both
$\vec{m}_{1,2}$ are oriented at $x$-direction. (e) All the conditions
are the same with (d) except the positions are the same with (c).
(f) Both the two dipoles move from the bottom to the top, with $\mathbf{r}_{1}=(0.5L_{x},\,0.5L_{y},\,d)$,
and the distance is fixed as $R=0.1L$, and $0<d<L_{z}$. Both $\vec{m}_{1,2}$
are oriented at $z$-direction. (g) All the conditions are the same
with (f) except $\vec{m}_{2}$ is oriented at $x$-direction. In (f,
g), the purple and blue lines indicate the contributions from the
images and mode-mediations respectively. The dipolar interactions
in freespace are plotted for comparison, namely, $V_{0}$ in Eq.\,(\ref{eq:classical})
(dot-dashed green), and $V(\omega)$ in Eq.\,(\ref{eq:V-ME}) (dashed
blue).}
\label{fig-interactions}
\end{figure*}

Based on the above discussions, now we consider the dipolar interaction
between two magnetic dipoles inside a rectangular 3D cavity (with
$L_{x,y,z}\equiv L$) given by Eq. (\ref{eq:V21}). Without loss of
generality, here we focus on the interaction term between two resonant
transition dipoles, namely, the interaction term $V_{2\leftarrow1}^{uv,ab}(\omega)\,\hat{\tau}_{2}^{uv}\hat{\tau}_{1}^{ab}$
as demonstrated in Fig. \ref{fig-demo}, so as to make a close comparison
with the previous freespace results (\ref{eq:classical}, \ref{eq:V-ME})
(the dipole frequency is set as $\omega\equiv20\,cL^{-1}$). Hereafter
we denote this interaction strength as $V_{2\leftarrow1}^{\text{cav}}$
for simplicity. The numerical results for different configurations
are shown in Fig.\,\ref{fig-interactions}.

In Fig.\,\ref{fig-interactions}(a, b), dipole-1 (red) is placed
at the center of the cavity, dipole-2 (green) moves from the center
to the boundary ($\mathbf{r}_{2}=\mathbf{r}_{1}+R\,\mathbf{e}_{x}$
with $0<R<L_{x}/2$), and both dipoles are oriented at the $z$-direction.
As mentioned above, the full interaction contains the contributions
from the field propagations in both way $V_{1\leftrightarrow2}^{\text{cav}}=(V_{2\leftarrow1}^{\text{cav}}+V_{1\leftarrow2}^{\text{cav}})/2$,
and clearly these two contributions $V_{2\leftarrow1}^{\text{cav}}(\omega)$
and $V_{1\leftarrow2}^{\text{cav}}(\omega)$ are not exactly the same
with each other {[}Fig.\,\ref{fig-interactions}(a, b){]}, which
is different from the above isotropic situation in freespace {[}Eq.
(\ref{eq:V21_uv,ab}){]}.

As a comparison, the dipolar interactions in freespace with the same
conditions are also presented {[}Eqs. (\ref{eq:classical}, \ref{eq:V-ME}),
see the green and blue lines{]}. In the regime $\omega R/c>1$, the
interaction $V_{2\leftarrow1}^{\text{cav}}$ exhibits significant
oscillations with the distance $R$, which is more drastic than the
freespace result $V(\omega)$ {[}see the dashed blue line and Eq.
(\ref{eq:V-ME}){]}. When the distance between the two dipoles is
quite small (in the regime $\omega R/c\ll1$), $V_{2\leftarrow1}^{\text{cav}}$
just returns the static dipolar interaction in freespace (\ref{eq:classical})
with the power law dependence $\sim R^{-3}$ {[}see the log-log scale
inset in Fig.\,\ref{fig-interactions}(a){]}. That means, when the
two dipoles are far from the conductor boundary, their interaction
well returns the freespace situation \citep{agarwal_microcavity-induced_1998},
and this is also consistent with the situation in classical electrodynamics. 

Accordingly, we consider the situation that the two dipoles are placed
near the conductor plane. In Fig.\,\ref{fig-interactions}(c), dipole-1
is set near the bottom center $\mathbf{r}_{1}=(0.5L_{x},\,0.5L_{y},\,0.01L_{z})$,
and still dipole-2 moves away from dipole-1 ($\mathbf{r}_{2}=\mathbf{r}_{1}+R\,\mathbf{e}_{x}$
with $0<R<L_{x}/2$). The dipole orientations are the same as Fig.\,\ref{fig-interactions}(a,
b). Again, in the short-distance regime, $V_{2\leftarrow1}^{\text{cav}}$
well returns freespace result (\ref{eq:classical}). But in the long-distance
regime, it turns out the interaction $V_{2\leftarrow1}^{\text{cav}}$
is significantly suppressed comparing with the freespace results (\ref{eq:classical},
\ref{eq:V-ME}). The reason is, the dipole field $\hat{\mathbf{B}}_{\text{d1}}(\mathbf{r})$
generated from dipole-1 should follow the boundary condition near
the conductor plane during its propagation, thus $\hat{\mathbf{B}}_{\text{d1}}(\mathbf{r})$
tends to be parallel with the conductor plane. Therefore, since here
dipole-2 is perpendicular to the conductor plane, their local interaction
$-\hat{\boldsymbol{\mathfrak{m}}}_{2}\cdot\hat{\mathbf{B}}_{\text{d1}}(\mathbf{r}_{2})$
tends to vanish to zero {[}Eq.\,(\ref{eq:V21}){]}.

Similar comparison is also made for two dipoles oriented in $x$-direction
{[}Fig.\,\ref{fig-interactions}(d, e){]}. Since here the two dipoles
are parallel to the conductor plane, there is no suppressing behavior
as Fig.\,\ref{fig-interactions}(c) when the dipoles are placed near
the cavity bottom, and the interactions $V_{2\leftarrow1}^{\text{cav}}$
in Fig.\,\ref{fig-interactions}(d, e) look similar to each other.

To see this mechanism more clearly, we consider the distance between
the two dipoles is fixed as $\mathbf{r}_{2}=\mathbf{r}_{1}+R\,\mathbf{e}_{x}$
with $R\equiv0.1L$, and they both move from the cavity bottom to
the top {[}$\mathbf{r}_{1}=(0.5L_{x},\,0.5L_{y},\,d)$ with $0<d<L_{z}$
Fig.\,\ref{fig-interactions}(f){]}, and clearly the interaction
$V_{2\leftarrow1}^{\text{cav}}$ approaches zero when the two dipoles
(in $z$-direction) approach the top and bottom boundaries. The contributions
from the dipole images and mode-mediation are also presented respectively
{[}from $G_{A1}^{\sigma}$ and $G_{A2}^{\sigma}$ in Eq.\,(\ref{eq:G1-G2}){]},
comparing with the freespace interaction strength, which is a constant
due to the fixed distance. But if dipole-2 is oriented in $x$-direction,
$V_{2\leftarrow1}^{\text{cav}}$ would remain nonzero at the boundaries
$d\rightarrow0,\,L_{z}$ {[}Fig.\,\ref{fig-interactions}(g){]}.

The influence from the conductor boundary also can be understood from
the demonstration in Fig.\,\ref{fig-demo}. When dipole-1 is placed
near the conductor plane, the dipole field felt by dipole-2 comes
from both dipole-1 and its image. In the bulk regime far from the
boundaries {[}see Fig.\,\ref{fig-interactions}(f, g){]}, the interaction
strength exhibits significant oscillations varying with positions,
which comes from the spatial distribution of the mediating modes,
especially the ones nearly resonant with the dipole frequency.

\section{Discussions}

By properly considering the propagation of the dipole field, we obtain
the full magnetic dipolar interaction which includes all the interaction
terms between permanent dipoles, resonant or non-resonant transition
dipoles, and even the counter-rotating interaction terms altogether.
The result applies for general boundary conditions, which already
have been enclosed in the tensor Green function $\tilde{\mathbb{G}}_{A}(\mathbf{r},\mathbf{r}';\omega)$,
and this is also consistent with the classical Maxwell equation. In
particular, we show the interaction for the dipolar interaction in
a rectangular 3D cavity, and how it is connected with previous results
under certain conditions.

From the above results, it is worth noticing that the dipolar interaction
exhibits significant dependence of three typical lengths, i.e., the
distance $R$ between the two dipoles, the distance $d$ between the
dipoles to the cavity boundaries, and the wavelength $\lambda$ of
the field modes nearly resonant with the dipole frequencies.

For example, the NV centers ($\omega_{\text{\textsc{nv}}}\simeq2.88\,\text{GHz}$)
in a nano-diamond interact with the $^{13}\text{C}$ nuclear spins
$(\omega_{\text{\textsc{c}}}\sim1\,\text{MHz})$ around through the
magnetic dipolar interaction \citep{doherty_nitrogen-vacancy_2013,wrachtrup_processing_2006,zhao_decoherence_2012}.
If the nano-diamond is placed in the center of a 3D cavity whose base
frequency is $\sim1\,\text{GHz}$ with the size $\sim10\,\text{cm}$
\citep{ball_loop-gap_2018}, the magnetic dipolar interaction between
the NV center and the nuclear spins around should be almost the same
as the static dipolar interaction (\ref{eq:classical}) in freespace,
since these magnetic dipoles are too far away from the cavity boundaries
{[}Fig.\,\ref{fig-interactions}(a){]}. If the nano-diamond is placed
quite near to the cavity boundary, or quite close to a metallic STM
tip, their dipolar interactions would be significantly changed.

On the other hand, considering some cold atoms are placed in an optical
cavity (usually $\sim100\,\mu\text{m}$), the distance $R$ between
the flying atoms, their distance to the cavity boundaries $d$, and
the cavity mode wavelength $\lambda$ would be comparable \citep{tang_tuning_2018,lushnikov_collapse_2002,griesmaier_comparing_2006,fattori_magnetic_2008,davis_photon-mediated_2019,devoe_observation_1996}.
In this case, the dipolar interaction in the cavity would have a complicated
position dependence as shown in Fig.\,\ref{fig-interactions}.

Throughout the discussion, the dynamics of the mediating dipole field
is simply described by the Maxwell equation and the corresponding
Green function. Thus, by properly changing to some other field equations,
this approach can be generalized to study the interaction mediated
by other kinds of fields, such as the exciton-polariton or phonon
field \citep{henriques_exciton-polariton_2021,espinoza_engineering_2021,huang_dipole-dipole_2012,cortes_super-coulombic_2017}.

In the above discussions about the cavity situation, we only focus
on the ideal conductors, and realistic situations may involve more
physical effects. For example, near the metal surface, the surface
plasmon induced by the electron density oscillation would influence
the EM field nearby, and that would bring in extra changes to our
above discussions \citep{yang_single_2020,henriques_exciton-polariton_2021}.
In principle, the interaction induced by these extra fields also can
be considered by the approach in this paper.

\vspace{0.8em}

\emph{Acknowledgments }- S.-W. Li appreciates quite much for the helpful
discussion with N. Wu, D. Xu, and B. Zhang in BIT. This study is supported
by NSF of China (Grant No. 11905007), Beijing Institute of Technology
Research Fund Program for Young Scholars.

\appendix

\section{The effective interaction mediated by one field mode \label{sec:The-mediated-interaction}}

We consider two dipoles both interact with one common field mode,
and the Hamiltonian of the three body system is $\hat{H}=\hat{H}_{0}+\hat{V}$,
where 
\begin{align}
\hat{H}_{0} & =\omega_{1}\hat{\tau}_{1}^{+}\hat{\tau}_{1}^{-}+\omega_{2}\hat{\tau}_{2}^{+}\hat{\tau}_{2}^{-}+\nu\hat{a}^{\dagger}\hat{a},\nonumber \\
\hat{V} & =g_{1}\hat{\tau}_{1}^{+}\hat{a}+g_{2}\hat{\tau}_{2}^{+}\hat{a}+\text{h.c.}
\end{align}
The Fr\"ohlish-Nakajima canonical transformation gives an effective
Hamiltonian \citep{frohlich_theory_1950,nakajima_perturbation_1955},
\begin{align}
\hat{H}_{\text{eff}} & =e^{-\hat{S}}\hat{H}e^{\hat{S}}=\hat{H}+[\hat{H},\hat{S}]+\frac{1}{2}\big[\hat{H},[\hat{H},\hat{S}]\big]+\dots\nonumber \\
 & \simeq\hat{H}_{0}+\frac{1}{2}[\hat{V},\hat{S}]
\end{align}
where the first order $\hat{V}+[\hat{H}_{0},\hat{S}]\equiv0$ is eliminated
by properly setting $\hat{S}:=A\hat{\tau}_{1}^{+}\hat{a}+B\hat{\tau}_{2}^{+}\hat{a}-\text{h.c.}$,
and that gives 
\begin{gather}
[\hat{H}_{0},\hat{S}]=(\omega_{1}-\nu)(A\hat{\tau}_{1}^{+}\hat{a}+\text{h.c.})+(\omega_{2}-\nu)(B\hat{\tau}_{2}^{+}\hat{a}+\text{h.c.}),\nonumber \\
\Rightarrow A=\frac{g_{1}}{\nu-\omega_{1}},\qquad B=\frac{g_{2}}{\nu-\omega_{2}}.
\end{gather}
Therefore, the effective interaction becomes
\begin{align}
\hat{V}_{\text{eff}} & =\frac{1}{2}[\hat{V},\hat{S}]=\big(\tilde{\beta}\hat{\tau}_{1}^{+}\hat{\tau}_{2}^{-}+\text{h.c.}\big)+\sum_{\alpha=1,2}\tilde{\xi}_{\alpha}\hat{\tau}_{\alpha}^{z}(\hat{a}^{\dagger}\hat{a}+\frac{1}{2})\nonumber \\
\tilde{\beta} & =\frac{1}{2}\big[\frac{g_{1}g_{2}^{*}}{\omega_{1}-\nu}+\frac{g_{2}g_{1}^{*}}{\omega_{2}-\nu}\big],\qquad\tilde{\xi}_{\alpha}=\frac{|g_{\alpha}|^{2}}{\omega_{\alpha}-\nu}.
\end{align}
The first term in $\hat{V}_{\text{eff}}$ eliminates the mediating
field mode and gives the interchange interaction between the two dipoles
as shown in Eq.\,(\ref{eq:adiabatic}), while the second term can
be regarded as a correction to the self-Hamiltonian $\omega_{\alpha}\hat{\tau}_{\alpha}^{+}\hat{\tau}_{\alpha}^{-}$
which depends on the mode state.

\section{The dynamical equation (\ref{eq:maxwell}) for the quantized magnetic
field\label{sec:The-dynamical}}

Here we show how Eq. (\ref{eq:maxwell}) for the quantized magnetic
field $\hat{\mathbf{B}}(\mathbf{r},t)$ is derived. We consider one
magnetic dipole at position $\mathbf{r}_{1}$ interacting with the
EM field, and the full dynamics of this system is described by $\hat{\mathcal{H}}=\hat{H}_{1}+\hat{V}_{1}+\hat{H}_{\text{\textsc{em}}}$,
where $\hat{H}_{1}$ is the self-Hamiltonian of the magnetic dipole,
and 
\begin{align}
\hat{H}_{\text{\textsc{em}}} & =\int d^{3}\mathbf{x}\,\big[\frac{1}{2}\epsilon_{0}\hat{\mathbf{E}}^{2}(\mathbf{x})+\frac{1}{2\mu_{0}}\hat{\mathbf{B}}^{2}(\mathbf{x})\big],\\
\hat{V}_{1} & =-\hat{\boldsymbol{\mathfrak{m}}}\cdot\hat{\mathbf{B}}(\mathbf{r}_{1})\nonumber 
\end{align}
are the Hamiltonian of the EM field, and the local interaction between
the magnetic dipole and the magnetic field respectively. Under the
Coulomb gauge, the quantized electric and magnetic field operators
read \citep{vogel_quantum_2006}
\begin{align}
\hat{\mathbf{E}}(\mathbf{r},t) & =\sum_{\mathbf{k}\varsigma}\sqrt{\frac{\hbar\omega_{k}}{2\epsilon_{0}V}}\mathbf{e}_{\mathbf{k}\varsigma}\big[ie^{i\mathbf{k}\cdot\mathbf{r}}\hat{a}_{\mathbf{k}\varsigma}(t)-\text{h.c.}\big],\\
\hat{\mathbf{B}}(\mathbf{r},t) & =\sum_{\mathbf{k}\varsigma}\sqrt{\frac{\hbar\omega_{k}}{2\epsilon_{0}V}}\frac{\mathbf{e}_{\mathbf{k}}\times\mathbf{e}_{\mathbf{k}\varsigma}}{c}\big[ie^{i\mathbf{k}\cdot\mathbf{r}}\hat{a}_{\mathbf{k}\varsigma}(t)-\text{h.c.}\big],\nonumber 
\end{align}
where $\mathbf{e}_{\mathbf{k}\varsigma}$ denotes the two polarization
directions perpendicular to the wave vector $\mathbf{k}$. These quantized
field operators follow the Heisenberg equation $\partial_{t}\hat{o}=\frac{1}{i\hbar}[\hat{o},\,\hat{{\cal H}}]$,
and that gives \begin{subequations} 
\begin{align}
\partial_{t}\hat{\mathbf{B}}(\mathbf{r}) & =\frac{1}{i\hbar}[\hat{\mathbf{B}}(\mathbf{r}),\,\hat{\mathcal{H}}]=-\nabla\times\hat{\mathbf{E}},\label{eq:dt B}\\
\partial_{t}\hat{\mathbf{E}}(\mathbf{r}) & =c^{2}\big[\nabla\times\hat{\mathbf{B}}-\mu_{0}\nabla\times\hat{\boldsymbol{\mathfrak{m}}}\delta(\mathbf{r}-\mathbf{r}_{1})\big].\label{eq:dt E}
\end{align}
\end{subequations} To obtain this result, the following commutation
relations are calculated,
\begin{align}
[ & \hat{\mathbf{E}}(\mathbf{r}),\,\hat{\boldsymbol{\mathfrak{m}}}\cdot\hat{\mathbf{B}}(\mathbf{r}_{1})]=\sum_{\mathbf{k}\varsigma,\mathbf{q}\sigma}\frac{-\hbar\sqrt{\omega_{k}\omega_{q}}}{2\epsilon_{0}Vc}\mathbf{e}_{\mathbf{k}\varsigma}(\hat{\boldsymbol{\mathfrak{m}}}\cdot\mathbf{e}_{\mathbf{q}}\times\mathbf{e}_{\mathbf{q}\sigma})\nonumber \\
 & \quad\cdot\big[e^{i\mathbf{k}\cdot\mathbf{r}}\hat{a}_{\mathbf{k}\varsigma}-e^{-i\mathbf{k}\cdot\mathbf{r}}\hat{a}_{\mathbf{k}\varsigma}^{\dagger},\,e^{i\mathbf{q}\cdot\mathbf{r}_{1}}\hat{a}_{\mathbf{q}\sigma}-e^{-i\mathbf{q}\cdot\mathbf{r}_{1}}\hat{a}_{\mathbf{q}\sigma}^{\dagger}\big]\nonumber \\
= & \sum_{\mathbf{k}\varsigma}\frac{-\hbar\omega_{k}}{2\epsilon_{0}Vc}(\hat{\boldsymbol{\mathfrak{m}}}\cdot\mathbf{e}_{\mathbf{k}}\times\mathbf{e}_{\mathbf{k}\varsigma})\mathbf{e}_{\mathbf{k}\varsigma}\,\big[e^{-i\mathbf{k}\cdot(\mathbf{r}-\mathbf{r}_{1})}-e^{i\mathbf{k}\cdot(\mathbf{r}-\mathbf{r}_{1})}\big]\nonumber \\
= & \sum_{\mathbf{k}}\frac{\hbar}{2\epsilon_{0}V}\,\mathbf{k}\times\hat{\boldsymbol{\mathfrak{m}}}\,\big[e^{-i\mathbf{k}\cdot(\mathbf{r}-\mathbf{r}_{1})}-e^{i\mathbf{k}\cdot(\mathbf{r}-\mathbf{r}_{1})}\big]\nonumber \\
= & \frac{i\hbar}{\epsilon_{0}}\nabla_{\mathbf{r}}\times[\hat{\boldsymbol{\mathfrak{m}}}\delta(\mathbf{r}-\mathbf{r}_{1})],
\end{align}
\begin{equation}
[\hat{\mathbf{E}}(\mathbf{r}),\,\hat{\mathbf{B}}^{2}(\mathbf{x})]=\frac{2i\hbar}{\epsilon_{0}}\nabla_{\mathbf{r}}\times\big[\hat{\mathbf{B}}(\mathbf{x})\delta(\mathbf{r}-\mathbf{x})\big].
\end{equation}
In the above calculations, we used the relations $\sum_{\varsigma=1,2}\mathbf{e}_{\mathbf{k}\varsigma}\mathbf{e}_{\mathbf{k}\varsigma}=\mathbf{1}-\mathbf{e}_{\mathbf{k}}\mathbf{e}_{\mathbf{k}}$,
and $\hat{\boldsymbol{\mathfrak{m}}}\cdot\mathbf{k}\times\mathbf{1}=-\mathbf{k}\times\hat{\boldsymbol{\mathfrak{m}}}$.

Notice that, Eqs. (\ref{eq:dt B}, \ref{eq:dt E}) and the current
term $\nabla\times[\hat{\boldsymbol{\mathfrak{m}}}\delta(\mathbf{r}-\mathbf{r}_{1})]:=\hat{\mathbf{J}}$
just have the same form as the classical Maxwell equations \citep{jackson_classical_1998}.
Taking the curl of Eq. (\ref{eq:dt E}) just gives the dynamical equation
(\ref{eq:maxwell}) for the quantized magnetic field $\hat{\mathbf{B}}(\mathbf{r},t)$
in the main text, which has the same form as the classical electrodynamics.
Indeed, during the canonical quantization of the EM field, the equations
of motion for the quantized operators should keep the same form as
their classical counterparts, which roots from the consistency from
the classical Poisson bracket to the quantum commutator $\{A,B\}\rightarrow\frac{1}{i\hbar}[\hat{A},\hat{B}]$.

\section{Eigen modes in the rectangular cavity\label{sec:Field-eigen-modes}}

Here we show the eigen modes of the cavity field given by $[\nabla^{2}+\mathbf{k}^{2}]\vec{\mathtt{A}}_{\mathbf{k}}(\mathbf{r})=0$
in the rectangular region $x,y,z\in[0,L_{x,y,z}]$. The boundary condition
requires $\hat{n}\times\vec{\mathtt{A}}_{\mathbf{k}}(\mathbf{r})=0$,
$\nabla\cdot\vec{\mathtt{A}}_{\mathbf{k}}(\mathbf{r})=0$ for $\mathbf{r}$
on the boundary planes. Denoting the vector eigen modes as $\vec{\mathtt{A}}_{\mathbf{k}}(\mathbf{r}):=(\mathtt{A}_{\mathrm{npq}}^{x},\,\mathtt{A}_{\mathrm{npq}}^{y},\,\mathtt{A}_{\mathrm{npq}}^{z})$,
the eigen modes reads \citep{park_accelerated_2009,sanamzadeh_fast_2019}
\begin{align*}
\mathtt{A}_{\mathrm{npq}}^{x}(\mathbf{r}) & =\sqrt{\frac{4(2-\delta_{\mathrm{n}0})}{V}}\,\cos\frac{\mathrm{n}\pi}{L_{x}}x\,\sin\frac{\mathrm{p}\pi}{L_{y}}y\,\sin\frac{\mathrm{q}\pi}{L_{z}}z,\\
\mathtt{A}_{\mathrm{npq}}^{y}(\mathbf{r}) & =\sqrt{\frac{4(2-\delta_{\mathrm{p}0})}{V}}\,\sin\frac{\mathrm{n}\pi}{L_{x}}x\,\cos\frac{\mathrm{p}\pi}{L_{y}}y\,\sin\frac{\mathrm{q}\pi}{L_{z}}z,\\
\mathtt{A}_{\mathrm{npq}}^{z}(\mathbf{r}) & =\sqrt{\frac{4(2-\delta_{\mathrm{q}0})}{V}}\,\sin\frac{\mathrm{n}\pi}{L_{x}}x\,\sin\frac{\mathrm{p}\pi}{L_{y}}y\,\cos\frac{\mathrm{q}\pi}{L_{z}}z,
\end{align*}
where $\mathbf{k}_{\mathrm{npq}}=(\mathrm{n}\pi/L_{x},\,\mathrm{p}\pi/L_{y},\,\mathrm{q}\pi/L_{z})$
and $\mathrm{n},\mathrm{p},\mathrm{q}\in\mathbb{Z}_{0}^{+}$. Here
$\mathtt{A}_{\mathrm{npq}}^{\sigma}(\mathbf{r})$ are normalized as
$\int_{V}\mathtt{A}_{\mathbf{k}}^{\sigma}(\mathbf{r})\,\mathtt{A}_{\mathbf{q}}^{\varsigma}(\mathbf{r})\,d^{3}r=\delta_{\sigma\varsigma}\delta_{\mathbf{kq}}$. 

\bibliographystyle{apsrev4-2}
\bibliography{Refs}
 
\end{document}